\documentclass[prl,twocolumn,preprintnumbers,amsmath,amssymb]{revtex4}

\usepackage{graphicx}

\def\w#1{$\omega_{#1}$}

\begin{document}
\title{The very slow expansion of an ultracold plasma formed in a seeded supersonic molecular beam of NO}

\author{J. P. Morrison}
\author{C. J. Rennick}
\author{E. R. Grant}
\altaffiliation{Author to whom correspondence should be addressed. Electronic mail:
edgrant@chem.ubc.ca}
\affiliation{Department of Chemistry, University of British Columbia, Vancouver, BC V6T 1Z3, Canada}

\date{\today}

\begin{abstract}

The double-resonant laser excitation of nitric oxide, cooled to 1 K in a seeded supersonic molecular beam, yields a gas of $\approx$10$^{12}$ molecules cm$^{-3}$ in a single selected Ryberg state.  This population evolves to produce prompt free electrons and a durable cold plasma of electrons and intact NO$^{+}$ ions.  This plasma travels with the molecular beam through a field free region to encounter a grid.  The atomic weight of the expansion gas controls the beam velocity and hence the flight time from the interaction region to the grid.  Monitoring electron production as the plasma traverses this grid measures its longitudinal width as a function of flight time.   Comparing these widths to the width of the laser beam that defines the initial size of the illuminated volume allows us to gauge the rate of expansion of the plasma.  We find that the plasma created from the evolution of a Rydberg gas of NO expands at a small but measurable rate, and that this rate of expansion accords with the Vlasov equations for an initial electron temperature of $T_{e} \approx 8~K$.

\end{abstract}


\maketitle

\section{introduction}

When a tunable pulsed laser promotes a substantial fraction of the atoms held in a magneto-optical trap (MOT) to a high Rydberg state (n $>$ 30), the ensemble spontaneously evolves on a microsecond timescale to produce an ultracold plasma \cite{Gallagher:2000,Dutta,Gallagher:2003,Li,Cummings}.  This phenomenon and the plasma that it forms have attracted a great deal of recent attention \cite{Killian_Science, Killian_Rost, Rolston_Physics}, because these rarified laboratory systems can display a degree of charged-particle correlation otherwise found only in thermonuclear explosions and the cores of dense stars.  

Coulomb interactions act to govern the dynamics of the charged particles in a plasma when their  electrostatic repulsion exceeds their thermal translational energy.  The point at which this occurs depends on the density and temperature of the plasma \cite{Ichimaru}.  The parameter, $\Gamma$, gauges the degree of correlation in terms of the dimensionless ratio,   

\begin{equation}
	\Gamma  = \frac{q^{2 }/4\pi \varepsilon _{0} a}{kT},
\label{equ:correlation}
\end{equation}

\noindent where $q$ is the charge, and $a$, the Wigner-Seitz radius, relates to the particle density, $\rho$, by, 

\begin{equation}
	4/3~\pi a^3  = 1/\rho.
\label{equ:wigner}
\end{equation}

\noindent Conditions under which $\Gamma$ substantially exceeds 1 can give rise to liquid-like or solid-like spatial correlations, leading ultimately to Coulomb crystallization and a quantum-state-detailed variant of the Mott insulator-to-metal phase transition \cite{vitrant}.  

Ultracold plasmas formed in MOTs at densities of the order of $10^{10}$ atoms cm$^{-3}$ have an initial Wigner-Seitz radius of $a= 3 ~\mu m$.  This average inter-particle spacing grows with a plasma expansion at rates that accord with electron temperatures falling from tens of $K$ to the order of $10~K$ \cite{Laha}.   At its maximum sometime during the 40~$\mu s$ observation window, these conditions give rise to electron correlation parameters in the range of $\Gamma_{e} \approx 0.1$.  

Recently, we have demonstrated the evolution to an ultracold plasma by NO molecules entrained in a supersonic expansion \cite{Morrison2008}.   This development holds interesting potential for the study of correlation in mesoscopic charged-particle ensembles.  The use of supersonic expansions substantially broadens the scope of systems that can be studied; any substance that can be volatilized~---~including large molecules and refractory metal clusters~---~can be entrained in a seeded supersonic molecular beam.  Heavy-particle internal degrees of freedom present new dimensions for plasma relaxation dynamics, including an accommodation of electron energy in ion degrees of freedom extending beyond the translational heat capacity to include rotational, and, in some cases vibrational and vibronic states.  

In our experiment, a seeded supersonic expansion provides NO at a local density of 5 x 10$^{13}$ molecules cm$^{-3}$ and a moving-frame translational temperature less than $1~K$.  Double-resonant laser excitation promotes about 10 percent of these molecules to a selected high-Rydberg state, from which they evolve to form a remarkably durable ultracold plasma.  At the Wigner-Seitz radius corresponding to this density ($a= 500 ~nm$), an electron temperature of $10~K$ would yield a correlation, $\Gamma_{e} \approx 4$.  

Our initial observations provide no direct means to quantify the ion or electron temperatures in the fully-formed plasma.  We do know, however, that the plasma expands very little during the 9~$\mu s$ observational window of those experiments.  

We have now more than tripled this observation time.  Replacing helium with neon, argon and krypton as expansion gases into which we seed NO, we have reduced the beam velocity to substantially increase the flight time in the field-free region of our apparatus.  By monitoring electron production as the plasma traverses the grid that defines the end of the field-free region, we measure the longitudinal width of this distribution of NO$^{+}$ ions and electrons as a function of flight times from 9 to 30 $\mu$s.  

Comparing these widths to the width of the laser beam that defines the initial size of the illuminated volume, we are now much better able to gauge the rate of expansion of the plasma.  We find that the plasma created in a molecular beam from a Rydberg gas of NO does expand at a small but measurable rate, and that this rate of expansion accords with the Vlasov equations for an initial electron temperature of $T_{e} \approx 8~K$.

\section{experimental}

A pulsed jet of NO, seeded nominally at 10 percent in He, Ne, Ar or Kr at a backing pressure of 5 atm, expands through the 0.5 mm diameter nozzle of a Series 9 pulse valve (Parker Hannifin).  Two centimeters downstream, the free jet passes through an electroformed Ni skimmer to enter a mu-metal shielded electron spectrometer as a 1 mm diameter differentially pumped supersonic molecular beam.  

\begin{figure}
	\includegraphics[width=\columnwidth]{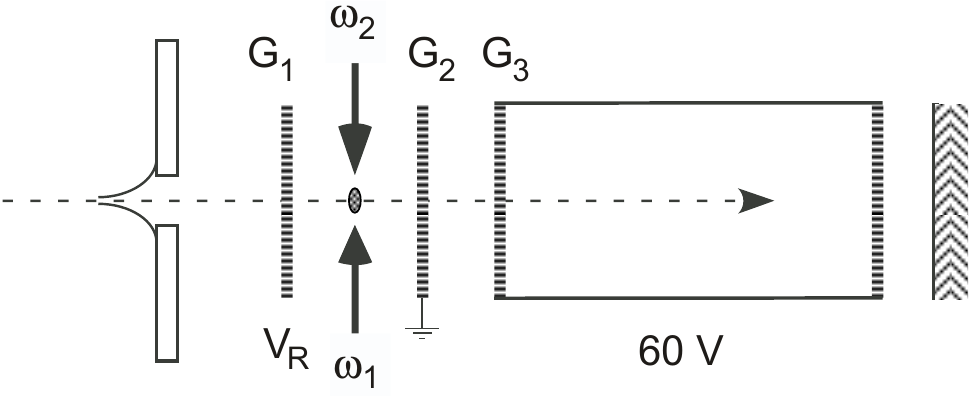}
	\caption{Schematic diagram showing the path of the molecular
	beam from a differentially pumped source chamber to the interaction region through a skimmer to enter a system of three grids ending in a flight tube capped by a microchannel plate (MCP) detector. Drawn to approximate scale with a G$_{1}$ to G$_{2}$ spacing of 2 cm.}
	\label{Figure_1}
\end{figure}

Figure~\ref{Figure_1} diagrams our two-stage electron spectrometer.  In the first stage, a pair of grounded grids, G$_{1}$ and G$_{2}$, held perpendicular to the axis of the molecular beam, define a field-free laser-molecular-beam interaction region.  Tunable pulses from two, frequency-doubled Nd:YAG-pumped dye lasers overlap to intersect the molecular beam halfway between the plates.  

Here, NO molecules absorb light in two resonant steps to reach a selected high Rydberg state.  These high-Rydberg NO molecules interact to form a plasma.  This plasma volume moves with the velocity of the molecular beam to traverse G$_{2}$, at which point it encounters a 60~V~cm$^{-1}$ field.  This field gradient extracts and accelerates plasma electrons to a multichannel plate detector.  On the timescale of our measurement, electrons extracted at G$_{2}$ appear instantaneously at the detector, and the profile of this signal thus integrates transverse slices of the electron density distribution of the plasma as it passes through G$_{2}$.   

\begin{figure}
	\includegraphics[totalheight=0.25\textheight]{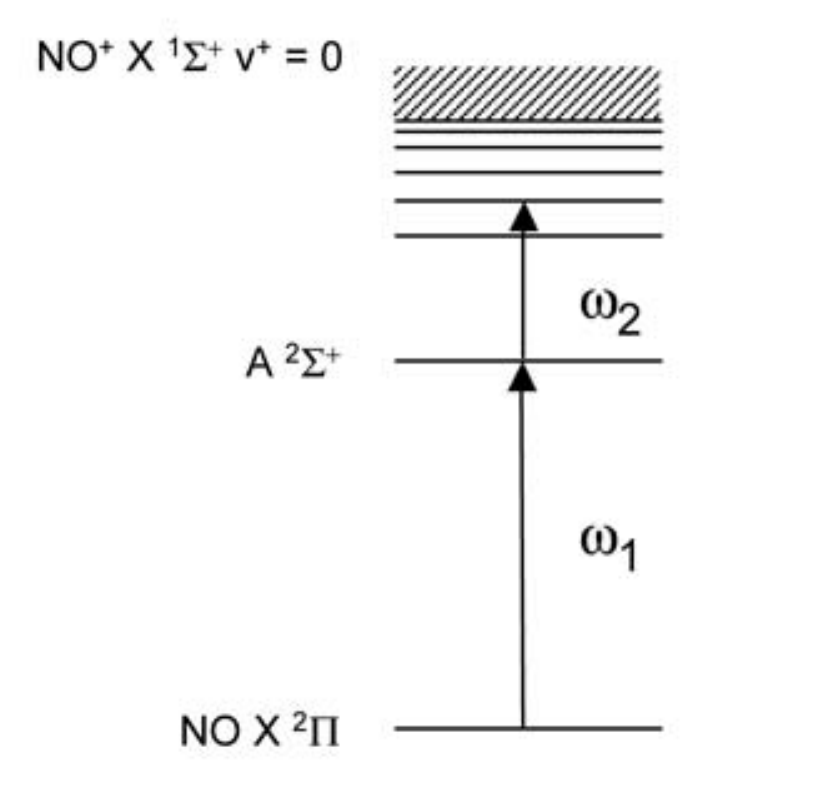}
	\caption{Excitation scheme for the population of $nf$ Rydberg states by double-resonant transitions via the $^{2}\Sigma^{+}, v=0, N=0,  J=\frac{1}{2}(+)$ state of NO.}
	\label{Figure_2}
\end{figure}

Figure~\ref{Figure_2} shows an energy diagram for the excitation of NO.  An initial laser pulse (\w1) pumps the transition from the ground,  X$^{2}\Pi$ state to rotational substates of the the $v=0$ vibrational level of the first electronically excited state, $^{2}\Sigma^{+}$.  A second laser (\w2), timed to coincide with \w1\, excites from a chosen level in this gateway system to a selected high Rydberg state situated below the lowest ionization threshold. 

At high laser pulse energies two-photon absorption of  \w1\  alone ionizes NO.  Monitoring the electron signal as a function of  the \w1\ wavelength under these conditions produces a rotationally resolved resonant ionization spectrum of the $v=0$ band of the $^{2}\Sigma^{+}$ state.  We analyze the intensities of lines in this spectrum to estimate the rotational temperature of NO under our expansion conditions.  

For plasma experiments, we attenuate \w1\  using a pair of Glan-Taylor polarizing prisms to eliminate any background signal from \w1 alone, and then counter propagate unfocussed \w1 and \w2 laser beams to intersect the molecular beam as indicated in Figure~\ref{Figure_1}.  This illumination geometry crosses a cylindrical photon field with a collimated molecular beam to produce a prolate ellipsoid excitation volume that propagates sideways with the velocity of the beam toward G$_{2}$.  The diameter of \w1\  at its intersection with the molecular beam determines the initial equatorial width of this ellipsoid.  

\begin{figure}
	\includegraphics[totalheight=0.13\textheight]{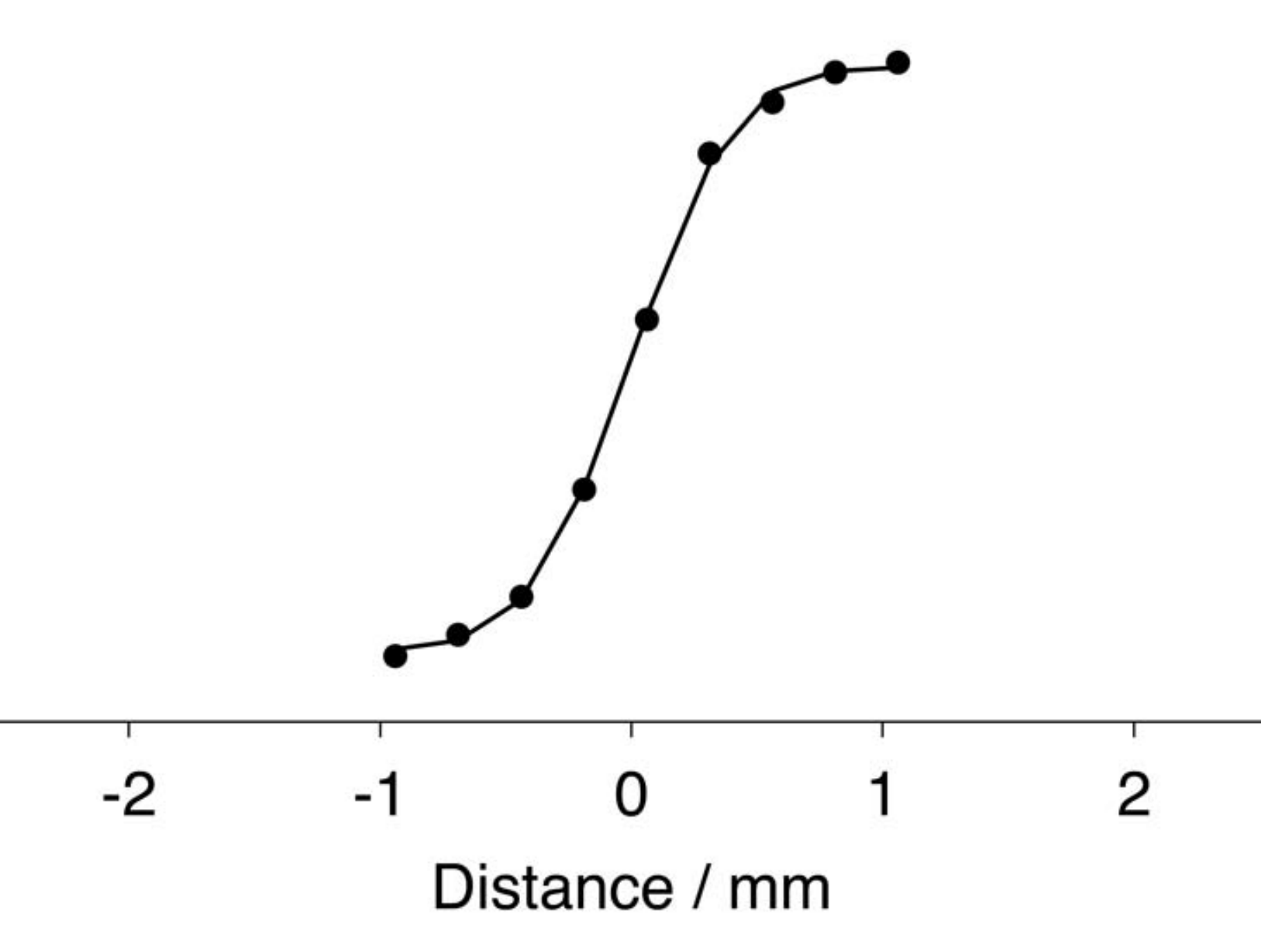}
	\caption{Horizontal profile of \w1\, the smaller of the two laser beam diameters, which determines the width of the plasma in the direction of molecular beam propagation .}
	\label{Figure_3}
\end{figure}

We experimentally gauge this limiting laser spot size by razor-blade tomography.  Figure~\ref{Figure_3} traces the integrated intensity of \w1\ as a function of the position of a razor blade scanned through the beam waist at a distance corresponding to the point at which the laser pulse crosses the axis of the molecular beam.  This measurement yields a value of 758 $\mu$m for the initial width of the active excitation volume along the propagation axis of the beam.  The divergence of the molecular beam after the skimmer determines a cross-beam length for this illuminated cylinder of approximately 5 mm.

\section{results}

\begin{figure}
	\includegraphics[totalheight=0.22\textheight]{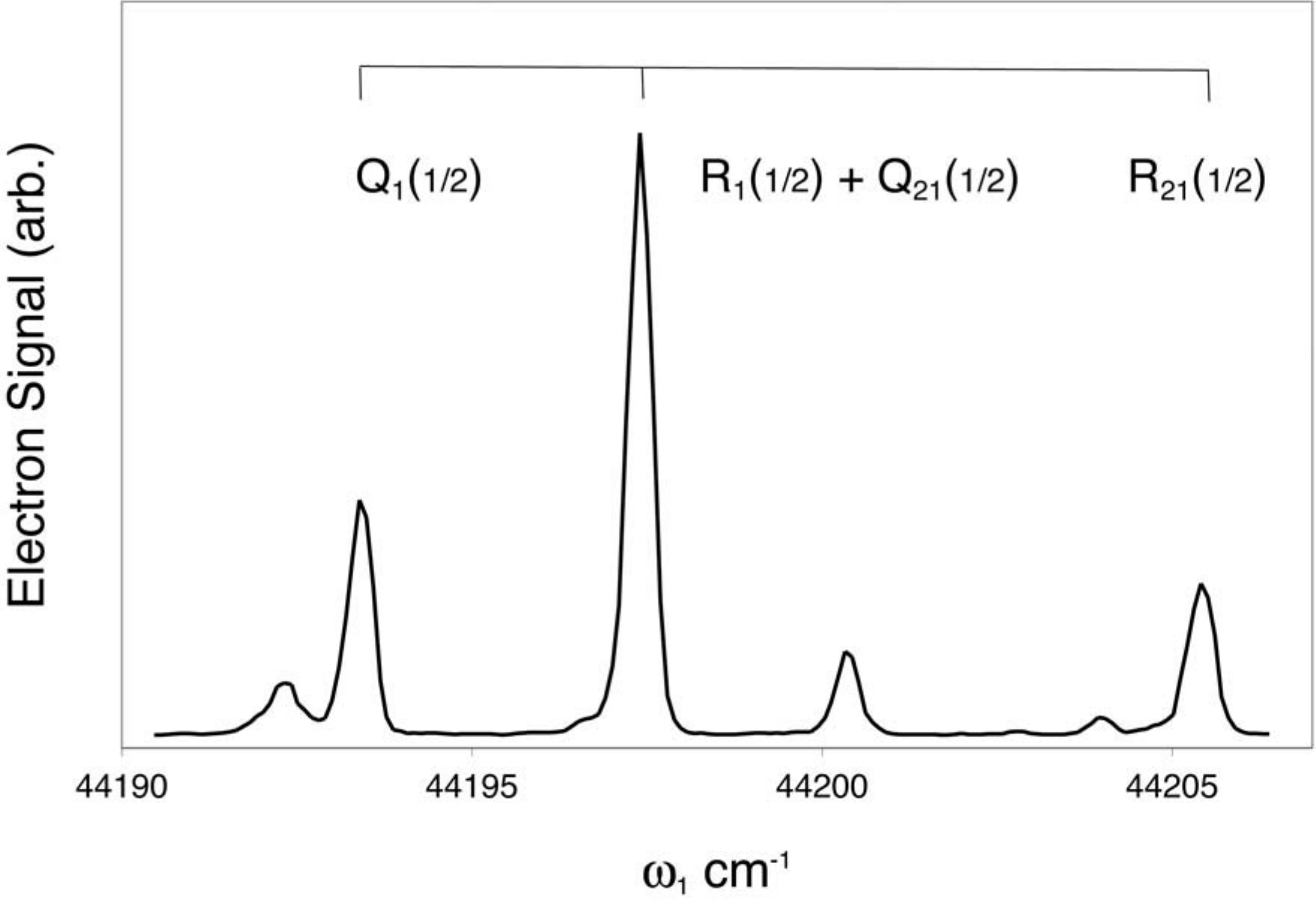}
	\caption{Resonant 1+1 ionization spectrum of the X $^{2}\Pi ~v'' = 0$) to A $^{2}\Sigma^{+} ~v' = 0$) transition in NO.  This pattern of intensities fits a ground state population distribution corresponding to a rotational temperature, $T^{\infty}_{R}$ = 2.5 K. }
	\label{Figure_4}
\end{figure}

A seeded supersonic expansion under the present conditions of backing pressure nozzle diameter cools a diatomic gas to a parallel temperature, $T^{\infty}_{\|}$, of about 700~mK, where $T^{\infty}_{\|}$ represents the second moment of the velocity distribution function along the axis of the molecular beam \cite{Miller}.  Supersonic expansion cools molecular rotation as well, but the cross section for rotational relaxation is smaller than that for translation.  As a result, when collisions cease, the terminal rotational temperature, $T^{\infty}_{R}$, exceeds $T^{\infty}_{\|}$.  Figure~\ref{Figure_4} shows a \w1\ scan of the resonant ionization spectrum of the X $^{2}\Pi ~v'' = 0$) to A $^{2}\Sigma^{+} ~v' = 0$) transition of NO under the conditions of our experiment, from which we estimate $T^{\infty}_{R} = 2.5 K$ .  

From the nozzle diameter and backing pressure, we estimate the centerline density of particles in the molecular beam to be $4.8\times10^{14}\ \mathrm{cm}^{-3}$  \cite{Morrison2008,Beijerinck}.  The seeding ratio of NO is 0.1.  At a rotational temperature of 3~K, 87 percent of these molecules populate the two parity components of the rotational ground state.  Two saturated steps of laser excitation transfer 12.5 percent of this population to a parity selected high Rydberg state.  In combination, these factors predict the density of excited NO available for plasma formation to be $5\times10^{12}\,\mathrm{cm}^{-3}$.  

\begin{figure}
	\includegraphics[totalheight=0.43\textheight]{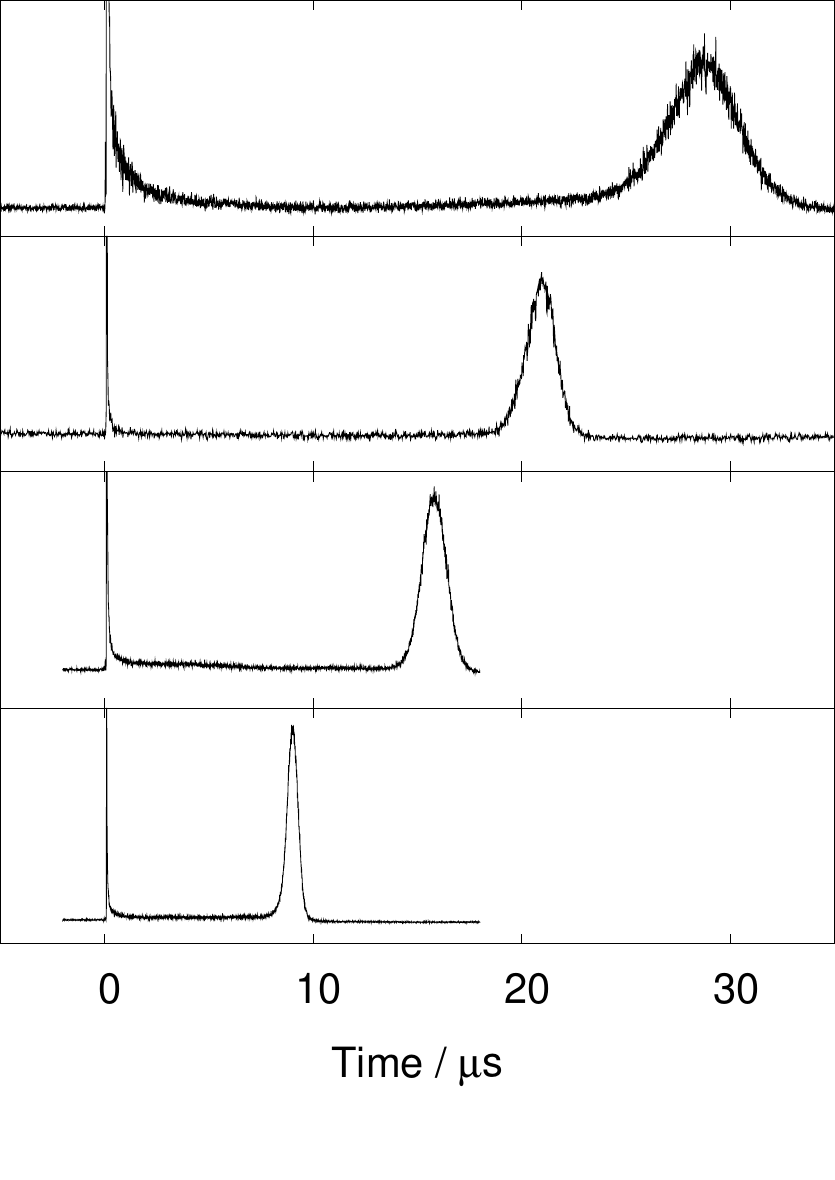}
	\caption{Electron signal waveforms appearing at G$_{2}$ following the substantial promotion of NO molecules to the $52f(2)$ Rydberg state in seeded supersonic expansions of He, Ne, Ar and Kr (from bottom to top) using the experimental flight path diagrammed in Figure~\ref{Figure_1}.}
	\label{Figure_5}
\end{figure}

To prepare a plasma, we reduce the power of \w1, set its frequency on the Q$_{1}(\frac{1}{2})$ line of this system (X $^{2}\Pi ~v'' = 0, N'' = 1, J'' = \frac{1}{2}(-)$ to $^{2}\Sigma^{+} ~v' = 0, N' = 0, J' = \frac{1}{2}(+)$), and tune \w2\ so that the total \w1\ + \w2\ energy approaches the adiabatic ionization threshold of NO.  Figure~\ref{Figure_5} shows electron signal waveforms obtained for expansions of NO in He, Ne and Ar, consisting of a prompt electron signal that appears within a few nanoseconds of \w2\ followed tens of microseconds later by a broader late signal.    

In previous work, we have established that these waveforms reflect the production of an ultracold plasma of electrons and molecular nitric oxide cations \cite{Morrison2008}.  The prompt signal arises from electrons released early in the formation of the plasma.  The late signal marks the passage of the quasi-neutral plasma through the grid, G$_{2}$ at the laboratory velocity of the molecular beam.  

\begin{figure}
	\includegraphics[totalheight=0.22\textheight]{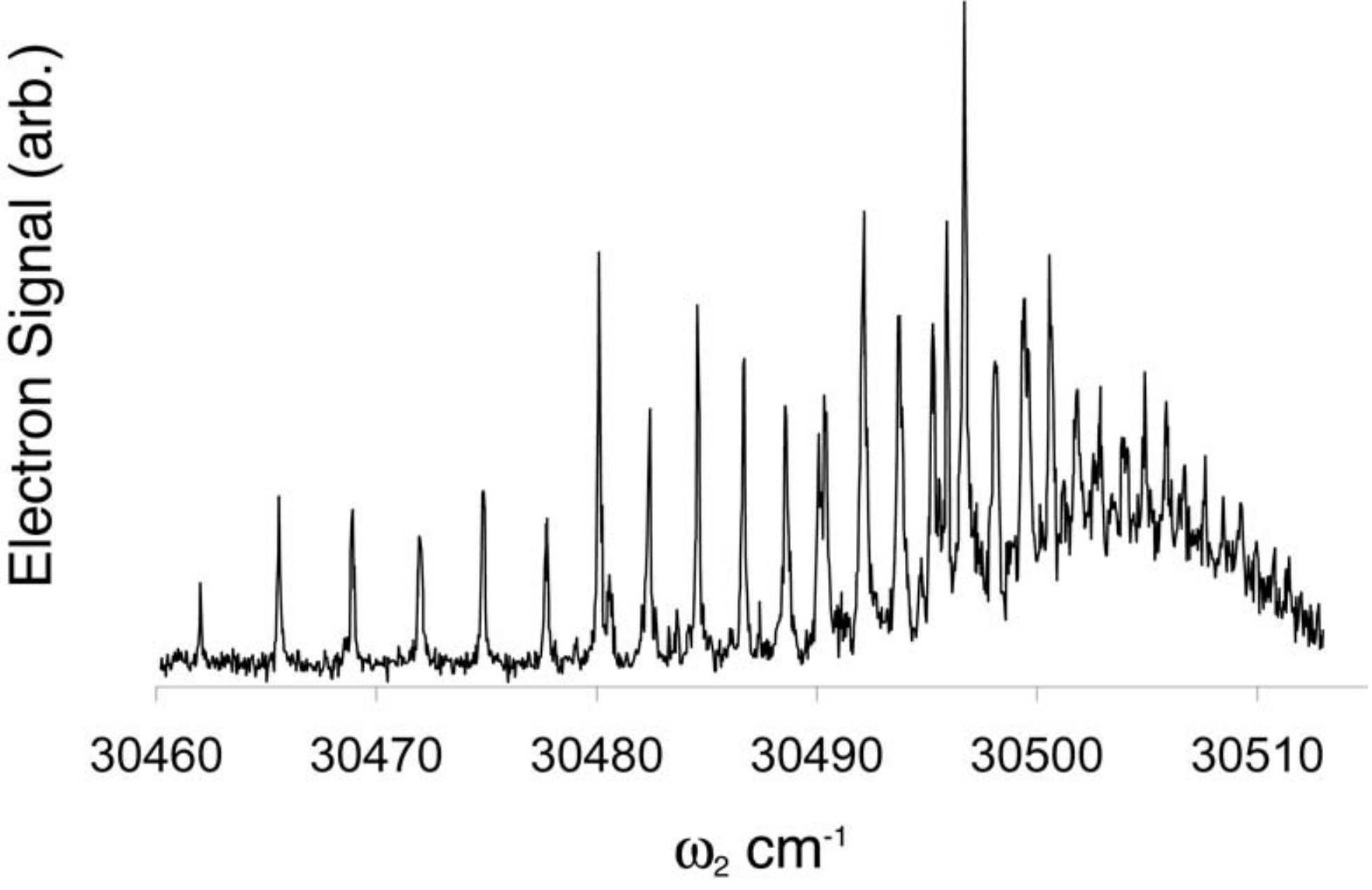}
	\caption{Resonances in the late signal observed scanning $\omega_{2}$. Spectrum assigned to lines in the $nf$ Rydberg series converging to the rotational level, $N^{+}=2$ in the vibrational ground state of NO$^{+}~X~^{1}\Sigma^{+}$. $52f(2)$ falls at an $\omega_{2}$ energy of 30,493.9~cm$^{-1}$, with a convergence limit of 30,534.5 cm$^{-1}$.}
	\label{Figure_6}
\end{figure}

Gating on the late signal and scanning the wavelength of \w2, we obtain a spectrum of closely spaced resonances, which is shown for a backing gas of Ar in Figure~\ref{Figure_6}.  We assign this spectrum to lines in the $nf(2)$ Rydberg series \cite{Vrakking}, where the number 2 refers to the rotational level of NO$^{+}$ to which the series converges.  The time traces shown in Figure~\ref{Figure_5} are obtained for \w2\ tuned to $52f(2)$.  

\begin{figure}
	\includegraphics[totalheight=0.38\textheight]{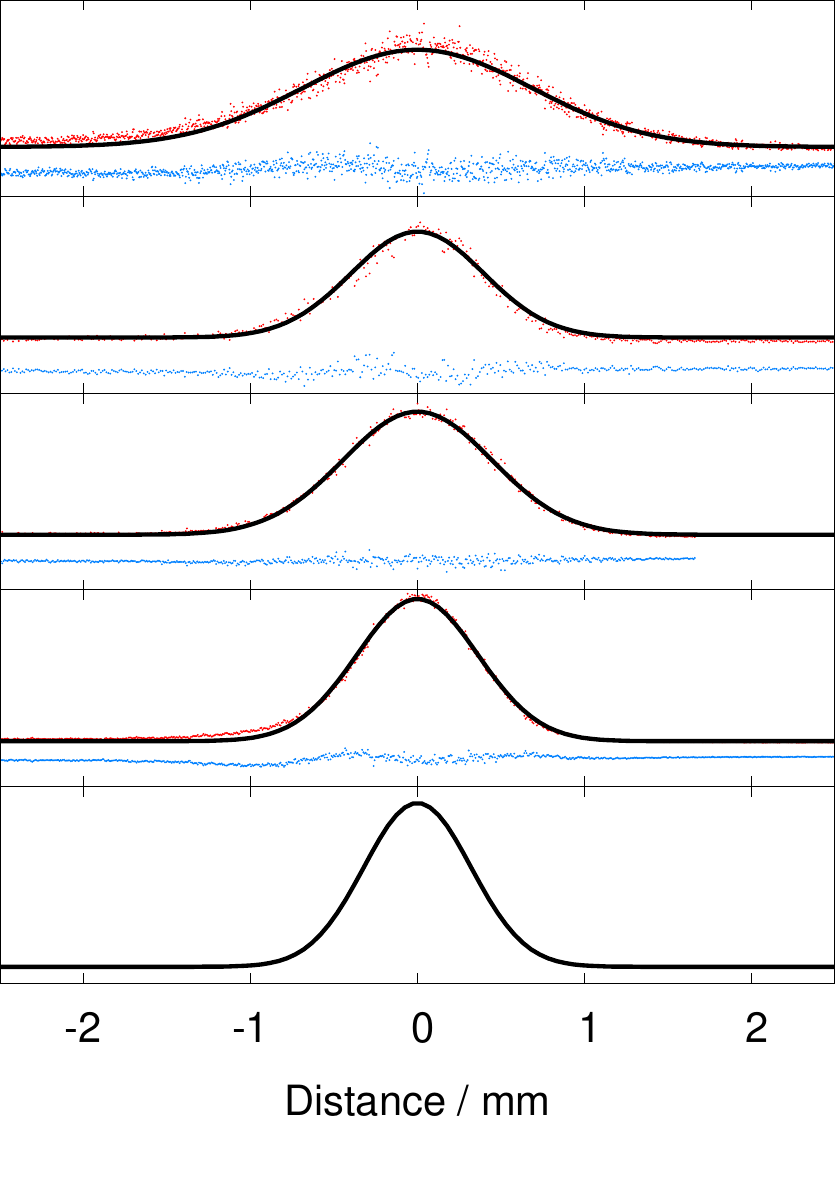}
	\caption{Late electron-signal temporal waveforms from Figure~\ref{Figure_5} transformed to yield spatial plasma distributions, displayed according to peak arrival time at G$_{2}$.  Solid curves show Gaussian fits with residuals below.  The bottom figure is derived from the measurement of laser width shown in Figure~\ref{Figure_3}, above which are plotted data for He, Ne. Ar and Kr carrier gases.}
	\label{Figure_7}
\end{figure}

For each expansion gas, the measured arrival time of the plasma at G$_{2}$, coupled with the known flight distance from the laser interaction region, yields a sharply defined laboratory-frame average velocitiy, $<v_{z}>$ by which to convert the observed temporal width to a spatial width along the propagation axis of the molecular beam.  Figure~\ref{Figure_7} plots the late-peak waveforms from Figure~\ref{Figure_5} transformed to distance.  We interpret these growing widths to represent the plasma density distribution as it expands in the spatial dimensions transverse to the laser illumination axis.  Distances measured in the z-dimension fit well with a representation of this expansion in terms of a two-dimensional Gaussian:

\begin{equation}
	n_{i}(y,z,t) = \frac{N_{i}}{\sigma(t)\sqrt{2\pi}} exp[\frac{-(y^{2}+z^{2})}{2\sigma^{2}(t)}]
\label{equ:gaussian}
\end{equation}

\noindent for which $\Gamma = 2\sqrt{2 ln2} \sigma(t)$ defines the full-width at half-maximum of the plasma density distribution in space, as defined by the time-varying standard deviation, $\sigma(t)$.  Table~\ref{Table_I} lists the arrival times and spatial widths of the electron-density images obtained in this way for all four expansion gases used in this study.

\begin{table}[h]
\begin{center}
\begin{tabular}{ccccc}\toprule
Carrier & Arrival Time& Velocity & $\sigma(t)$ & Width \\
Gas & ($\mu$s) &  (m s$^{-1}$) &   ($\mu$m)  &  ($\mu$m) \\
\hline
He & 9.0& 1358 & 364 & 856 \\

Ne & 15.8 & 758 & 446 & 1050 \\

Ar & 20.1 & 560 & 548 & 1291 \\

Kr & 28.7 & 394 & 686 & 1617 \\
\hline
\end{tabular}
\caption{\label{Table_I}Arrival times, estimated velocities and widths obtained by fitting Gaussian distributions to the spatial waveforms pictured in Figure~\ref{Figure_7}, as transformed from the flight-time dependent electron signals shown in Figure~\ref{Figure_5}.}
\end{center}
\end{table}

\section{Discussion}

By choice of the expansion gas, we systematically vary the laboratory velocity of nitric oxide in a differentially pumped seeded supersonic molecular beam.  Crossing this beam with the output of two dye lasers, we use double-resonant excitation to transform a prolate volume element with a minor axis diameter smaller than 1 mm into a dense distribution of NO molecules excited to a selected high-Rydberg state substantially below the lowest ionization threshold.   

This population of high-Rydberg molecules evolves to form a plasma of NO$^{+}$ cations and electrons.  The volume element containing this plasma moves through a field-free region with the propagation velocity of the molecular beam to traverse a perpendicular grid and enter a 60 V/cm field gradient.  Transmission of this volume element through the grid disrupts the plasma to produce a waveform of promptly detected electrons that maps the plasma charge-density distribution along the axis of the molecular beam.  

\begin{figure}
	\includegraphics[width=\columnwidth]{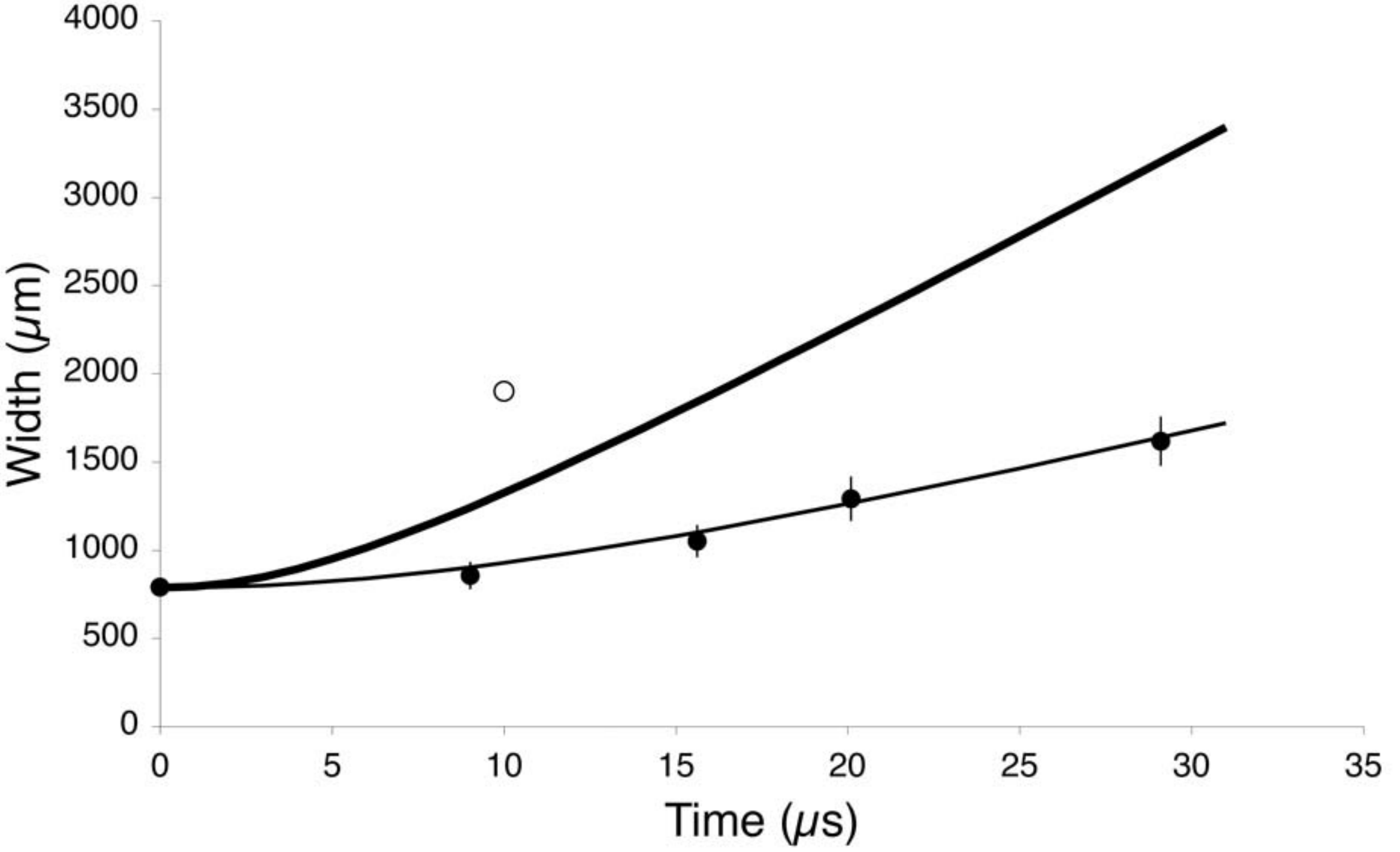}
	\caption{Plot of plasma FWHM at G$_{2}$ as a function of flight time, compared with width predicted after 10 $\mu$s for ambipolar expansion at $T_{e}$ = 40 K (open dot) and Vlasov expansion of a spherical Gaussian plasma for $T_{e}(0)$ = 40 K (bold line).  The line through the data represents a Vlasov fit for $T_{e}(0)$ = 7.8~K.  Error bars reflect the experimental uncertainty in the measured flight distance to G$_{2}$, which exceeds the error introduced by using velocity averages to transform time to distance. }
	\label{Figure_8}
\end{figure}

Converting these temporal images to displacements in space, we measure the physical width of the plasma along its laboratory propagation axis as a function of flight time from 9 to 30 $\mu$s.  Figure~\ref{Figure_8} plots these results.  With reference to the diameter of the initial illumination volume, we find that in 30 $\mu$s this measure of the plasma width approximately doubles.

The expansion rate of an ultracold plasma relates principally to the thermal energy of its electrons.  Considerable effort in the field of atomic ultracold plasma research has focused on the determination of the electron temperature over the life of the plasma and its relation to  plasma expansion \cite{RobicheauxandHanson,Roberts, Gupta, Fletcher, Laha}.  These measurements have generally supported the idea that, for plasmas prepared above threshold, the quantity,  $T_{e} = 2E_{e}/3k_{B}$, reasonably describes the initial electron temperature, so long as $E_{e}$, the excess energy, exceeds $\sim$ 50 cm$^{-1}$.  For plasmas produced by photoexcitation tuned lower in energy to release free electrons with less translational energy, and lower still to populate high-Rydberg states, electron heating mechanisms dominate to maintain initial electron temperatures in the range of 40 K.  
 
Electrons released by plasma formation expand much more rapidly than the colder (T$_{i} \approx$ 1 K), heavier positive ions.  However, the attractive force of a developing space charge acts to constrain the expanding electron spatial distribution to grow no faster than the timescale for cation motion.  By this Coulomb coupling, the thermal energy of the electrons drives overall plasma expansion, $viz$:  

\begin{equation}
	\frac{1}{2}M_{i}v^{2} = \frac{1}{2}k_{B}T_{e}
	\label{equ:expansion}
\end{equation}

For NO$^{+}$ in a plasma with an electron temperature of 40 K, this expression predicts an expansion rate of 100 $\mu$m $\mu$s$^{-1}$.  An open dot on Figure~\ref{Figure_8} shows the width after 10 $\mu$s predicted by Eq.~\ref{equ:expansion} for a plasma with an initial diameter, $\sigma(0)$ = 758 $\mu$m.  Clearly, the size expected for ambipolar expansion at  $T_{e}(0)$ = 40 K substantially exceeds our observations.  But, we should expect this simple picture to overestimate the expansion rate.  By expanding, the electron charge distribution does work on the cations, so the temperature of the system, and thus the force driving its expansion, should fall with time.  

A fuller account of the coupled evolution of the particle density and energy distribution functions is provided by the Vlasov equations \cite{Manfredi}, which form the foundation of the kinetic theory of plasmas, and have analytical forms for many types of low-density collisionless plasmas \cite{Dorozhkina}. Laha and coworkers \cite{Laha} have applied an analytical self-similar solution to describe the expansion of a quasi-neutral ultracold plasma configured to have a spherically symmetric, Gaussian ion-density distribution.  The evolution timescale for such a plasma is given by:

\begin{equation}
	\tau_{\mathrm{exp}} = \sqrt{\frac{m_i \sigma(0)^{2}} {k_{B}\left[T_{e}(0)+T_{i}(0)\right]}}\
~\mathrm{,}\\
	\label{equ:tauexp}
\end{equation}

\noindent where the quantities denoted $T_{\alpha}(0)$ refer to the initial temperature
of the electrons, $T_{e}(0)$, and the ions, $T_{i}(0)$.   $m_{i}$ is the mass of the ions.  

The electron and ion temperatures fall with time according to 

\begin{equation}
	T_\alpha(t) = \frac{T_\alpha(0)}{1 + \frac{t^2}{\tau_\mathrm{exp}^2}}
	\label{equ:temp}~\mathrm{,}
\end{equation}

\noindent and the diameter of the plasma expands as

\begin{equation}
	\sigma(t) = \sqrt{\sigma(0)^2\left[1 + 
	\frac{t^2}{\tau_\mathrm{exp}^2}\right]}\,~\mathrm{.}
	\label{equ:diam}
\end{equation}

This simple formalism conforms very well with experimental measurements of ion positions and correlated expansion velocities in quasi-neutral ultracold plasmas of strontium prepared to have electron temperatures in the range of T$_{e}$(0) = 40 K and above \cite{Laha}.  

For reference, the bold line on Figure~\ref{Figure_8} gives the time-dependent width predicted by Eq.~\ref{equ:diam} for a plasma with the ion mass of NO, an initial width of 785 $\mu$m and initial electron and ion temperatures of $T_{e}(0)$ = 40 K and $T_{i}(0)$ = 1 K.  The curve shown significantly overestimates the rate at which our plasma expands.  

The collimated laser-crossed molecular-beam excitation geometry of our experiment produces a prolate ellipsoidal plasma volume, as opposed to a Gaussian sphere.  Nevertheless, tomography shows that our laser intensity distribution in the $yz$~plane is Gaussian, and slice measurements of the electron signal along the molecular beam propagation axis show that this distribution of excitation intensity gives rise to a Gaussian plasma density distribution in the equatorial dimension.  

Experiments by Cummings $et~al.$ \cite{Cummings} form an ellipsoidal plasma volume of similar aspect ratio in a calcium MOT.  They determine an equatorial expansion rate by modeling the time-dependent Ca$^{+}$ laser-induced fluorescence signal.  Though the analytical self-similar solution of the Vlasov equations for a spherically symmetric Gaussian plasma do not extend to this geometry, the expansion rate deduced by Cummings $et~al.$ \cite{Cummings} for calcium accords almost exactly with that found experimentally for the spherical strontium plasma of Laha et al.  \cite{Laha} when matched for initial electron energy and scaled to account for the ion mass difference.  

This suggests that a simple fit of the spherical formalism to the rate of expansion in the equatorial dimension provides a reasonable gauge of electron temperature for Gaussian ellipsoidal plasma geometries.  

The line drawn through the data on figure~\ref{Figure_8} represents a Vlasov fit to the equatorial plasma diameters measured in our experiment, returning initial electron temperature of $T_{e}(0)$ = 7.8~K (for an assumed ion temperature of $T_{i}(0)$ = 1~K).  Remarkably, by this fit, the Vlasov equations hold that after 30 $\mu$s the electron temperature of our plasma has fallen to 1.6 K.  By this point, plasma expansion has doubled the Wigner-Seitz radius, but the predicted electron correlation remains substantial, $\Gamma_{e}~\approx$~10.  

This plasma forms following nitric oxide Rydberg-Rydberg Penning ionization collisions.  Some initially formed electrons leave the excitation volume immediately, creating an electrostatic trap in which an electron-NO* avalanche ionizes the remaining Rydberg molecules.  Pillet, Pohl and coworkers have recognized the collisional ionization of Rydberg atoms as a cooling mechanism for electrons in ultracold plasmas \cite{Pohl:2006}.  MOT experiments appear to show that the properties of atomic ultracold plasmas vary continuously with excitation energy tuned across the ionization threshold (c.f. Figure 5 in reference \cite{Cummings}).  Thus, we expect the plasma formed from Rydberg NO molecules to exhibit a relatively low electron temperature.  

However, as mentioned above, efficient heating mechanisms appear to elevate electron temperatures, even in systems prepared very near threshold, to levels much higher than seem to be evident here.  Chief among these for the lowest initial values of T$_{e}$ is three-body recombination \cite{Fletcher}.  

In simple physical terms, three-body recombination occurs when an electron approaches an ion to within a critical radius, $r_{T}$, during which time a collision with a second electron carries away the binding energy of the pair.  By this mechanism, any ion-electron collision resonance bound at least by the energy to which the electron collisionally thermalizes can remain bound.  Thus, T$_{e}$ determines the largest orbital radius stabilized by three-body recombination (the Thomson radius \cite{Bates}) by the energy balance:

\begin{equation}
	\frac{3}{2} k_{B} T_{e}=\frac{e^{2}}{4\pi \epsilon_{0} r_{T}}
	\label{equ:rt}
\end{equation}

This largest, lowest binding energy orbital dominates recombination by virtue of the cross sections for electron capture and collisional thermalization, which scale as $r^{2}$.  Thus, the rate of three-body recombination vastly accelerates as the electron temperature approaches 0~K.  Various collision theory formulations developed to model three-body recombination consistently yield third order rate constants that scale with temperature as $T_{e}^{-4.5}$ \cite{ Makin, Bates, Mansbach,Flannery, RobicheauxandHanson, Kuzmin}.  Calculations for ultracold atomic plasmas show that three-body recombination makes it difficult to form a plasma with an electron temperature less than 25 K \cite{RobicheauxandHanson}. 

The very high rates predicted for low-temperature three-body recombination follow directly from the very large values of $r_{T}$ sampled in theory by low-energy recombining and deactivating electrons, or equivalently, the very high-values of maximum principal quantum number, $n_{max}$ (and accompanying orbital degeneracy), set by temperature in detailed-balance integrals.  

However, conventional formulations of the three-body recombination rate neglect the effect of plasma density on the accessible scale of $r_{T}$ or $n_{max}$.  For an electron temperature, $T_{e}$ = 2 K, the average kinetic energy, 2.6~meV, dictates a Thomson radius of 560 nm.  For a typical MOT plasma with a density of $3 \times 10^{15}~m^{-1}$, this lies well within the Wigner-Seitz radius of 4 $\mu$m.  By contrast, we estimate the plasma formed under our molecular beam conditions to have an ion density of $2 \times 10^{18}~m^{-3}$, or a Wigner-Seitz radius of 500 nm.  Models incorporating the effects of ion density on high-$n$ capture yield predicted rate coefficients that scale as $T_{e}^{-1}$, significantly reducing the predicted rate of three-body recombination \cite{Hahn1, Hahn2}.  

Thus, it appears that conditions are right in our beam experiment to suppress three-body recombination and its attendant electron heating effects, diminishing the rate of plasma expansion in comparison with atomic systems prepared in MOTs.  

A suppression of ion-electron recombination seems also apparent in the absence of dissociative recombination.  We have noted previously \cite{Morrison2008} that a plasma lifetime of 9 $\mu$s exceeds expectations based on known cross sections for the dissociative recombination of NO$^{+}$\cite{Carata}.  Present results include NO$^{+}$ plasmas that maintain a high degree of integrity for as long as 30 $\mu$s.  

\section{Acknowledgements}

It is a pleasure to acknowledge helpful discussions with T. Gallagher, D. Luckhaus, T. Pohl and J. M. Rost.  This work was supported by the Natural Sciences and Engineering Research Council of Canada (NSERC), the Canada Foundation for Innovation (CFI) and the British Columbia Knowledge Development Fund (BCKDF).

 \bibliography{PRA-references}

\end{document}